\documentclass{PoS}
\usepackage{xspace}
\usepackage{ifthen} 
\newboolean{uprightparticles}
\setboolean{uprightparticles}{false} 
\usepackage{amssymb}
\usepackage{amsfonts}
\usepackage{upgreek}
\usepackage{lhcb-symbols-def}
\usepackage{overpic}
\usepackage{lineno}

\title{Measurements of \CP violation in \B mixing through $\B \to \jpsi X$ decays at LHCb}

\ShortTitle{\CP violation in $\B \to \jpsi X$ decays}

\author{\speaker{Greig A. Cowan\footnote{On behalf of the LHCb collaboration.}}\\
        University of Edinburgh\\
        E-mail: \email{g.cowan@ed.ac.uk}}

\abstract{\B mesons provide an ideal laboratory for measurements of \CP violation and searches for \CP violation
beyond the Standard Model. Recent measurements of the mixing phases of the \Bs and \Bd mesons, \phis\
and $\sin2\beta$, using decays to $J/\psi X$ final states are presented. In view of future improved measurements,
a good understanding of pollution from sub-leading penguin topologies in these decays is needed.
Those can be probed using suppressed decays like $\Bs \to J/\psi\KS$ and $\Bs \to J/\psi \overline{K}^{*0}$.
Recent results using these decay modes are presented.}

\FullConference{The European Physical Society Conference on High Energy Physics\\
		22--29 July 2015\\
		Vienna, Austria}

\begin{document}

\section{\CP violation in \B mixing and decay}
\label{sec:intro}
\begin{figure}[t]
\centering{
\includegraphics[scale=0.13]{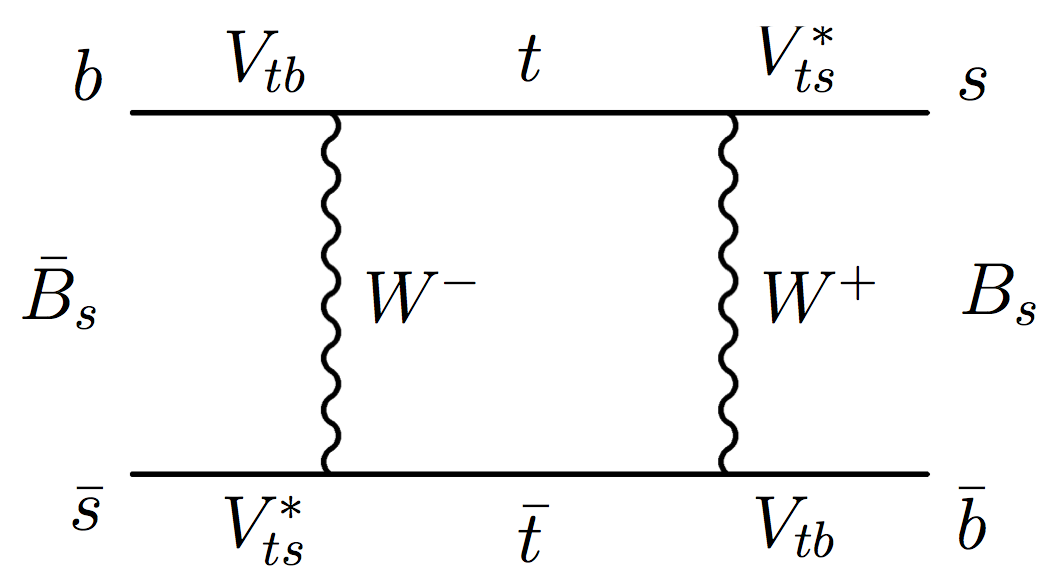}
\includegraphics[scale=0.67, clip=true, trim=40mm 185mm 30mm 56mm]{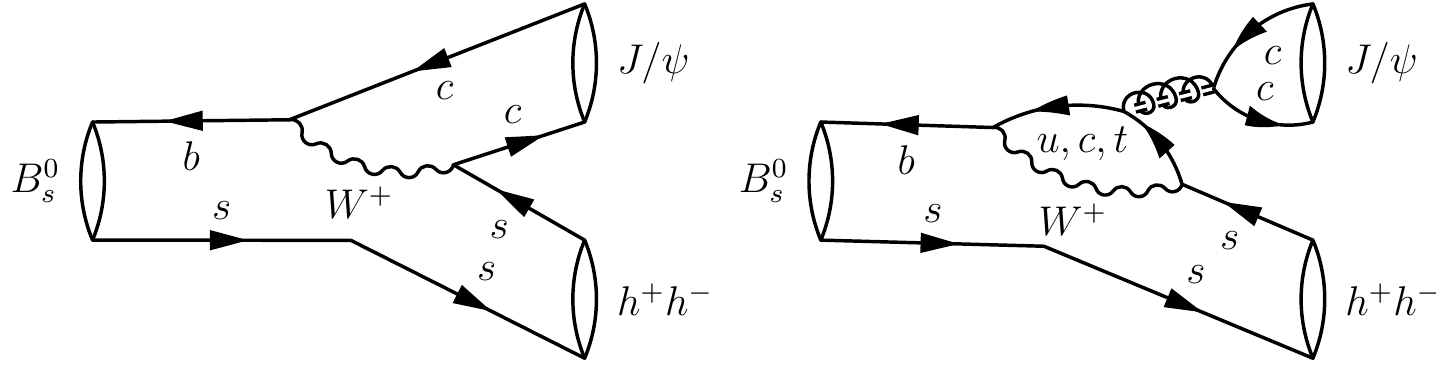}
}
\caption{Feynman diagrams showing $\Bs$ meson mixing (left) and example $b\to c\overline{c}s$ transitions
via a tree-level (middle) and penguin process (right).
\label{fig:feynman}}
\end{figure}

In both the \Bd and \Bs meson systems a \CP violating phase $\phi_{d,s}$
arises through the interference of $B^0_{(s)}$ mesons
decaying via $b\to c\overline{c}s$ transitions to \CP eigenstates and those
decaying after oscillation\footnote{Charge-conjugation is implicit unless stated otherwise.}.
Figure~\ref{fig:feynman} shows examples of these mixing and decay processes
for the \Bs system.
In the Standard Model (SM), ignoring sub-leading penguin processes, $\phis$ is equal
to $-2\beta_s$, where $\beta_s \equiv arg[-(V_{ts}V^*_{tb})/(V_{cs}V^*_{cb})]$ with $V_{ij}$ being elements of the
quark-mixing matrix~\cite{Kobayashi:1973fv,Cabibbo:1963yz}. Similarly, in the \Bd system the $\phi_d$
is equal to  $2\beta$, where $\beta \equiv arg[-(V_{cd}V^*_{cb})/(V_{td}V^*_{tb})]$.
Global fits to experimental data give precise determinations:
$\phis = -0.0365\pm 0.0012\rad$ and $\sin 2\beta = 0.771^{+0.017}_{-0.041}$~\cite{Charles:2015gya}.
These phases could be modified if non-SM particles contribute to the
\B meson oscillation~\cite{Buras:2009if, Chiang:2009ev} and, therefore, are the
subject of many experimental measurements. Typically the
measurements require the study of the \CP asymmetry as a function of the \B meson
decay time, $t$. The asymmetry is defined as,
 \begin{equation}
 A_{\CP}(t) \equiv \frac{\Gamma_{\overline{B}^0\to f} - \Gamma_{\Bd\to f}}{\Gamma_{\overline{B}^0\to f} + \Gamma_{\Bd\to f}}
 =
  \frac{S_f \sin(\dm\, t) - C_f \cos(\dm\, t)}
  {\cosh(\Delta\Gamma\, t/2) + A_{\Delta\Gamma} \sinh(\Delta\Gamma\, t/2)},
  \label{eqn:asymm}
 \end{equation}
where $\Gamma_{\Bd\to f}$ is the rate of the $\Bd \to f$ decay, $C_f \equiv \frac{1-|\lambda_f|^2}{1+|\lambda_f|^2}$,
$S_f \equiv \frac{2{\rm Im}(\lambda_f)}{1+|\lambda_f|^2}$ and $A_{\Delta\Gamma} \equiv -\frac{2{\rm Re}(\lambda_f)}{1+|\lambda_f|^2}$.
Here, $\Delta m$ is the oscillation frequency of the \B meson system, $\Delta\Gamma$ is the
width difference between the light and heavy eigenstates of the \B system. The parameter
$\lambda_f \equiv \frac{q}{p}\frac{\overline{A}_f}{A_f}$ describes \CP violation in the interference
between mixing and decay, with the \CP violating phase defined by $\phi \equiv -\arg(\lambda_f)$.

These proceedings will first review the experimental measurements
of the \CP violating phases, focussing on recent analyses of $B\to\jpsi X$ decays
($\jpsi\to\mu^+\mu^-$) from the LHCb collaboration
that have been obtained using $3.0\invfb$ of $pp$ collision data recorded at a
centre-of-mass energies of $\sqrt{s} = 7$ and $8\tev$ at the Large Hadron Collider.
As will become clear, the increasing precision of these measurements now make it essential
to test the assumption that $b\to c\overline{c}s$ penguin processes are
small and therefore have little contribution to the \CP violating phases. As such, the second half of the
proceedings will focus on the latest experimental developments in this area.

\section{$\sin2\beta$ from $\Bd\to\jpsi\KS$}

\begin{figure}[t]
\centering{
\includegraphics[scale=0.9]{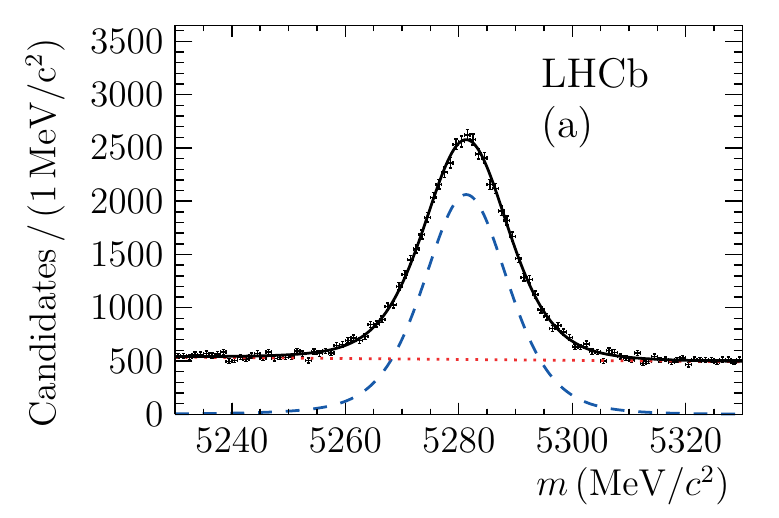}
\begin{overpic}[scale=0.9]{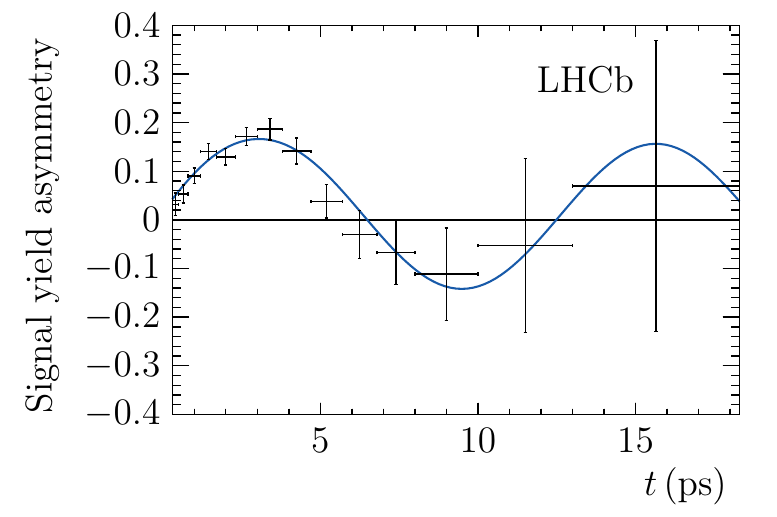}
 \put (30,55) {(b)}
\end{overpic}
}
\caption{(a) Distribution of the $\jpsi\KS$ invariant mass for $\Bd\to\jpsi\KS$ candidates with the signal (blue-dotted line)
and background (red-dotted) components of the maximum likelihood fit shown. 
(b) The signal asymmetry as a function of \Bd decay time.
\label{fig:jpsiKS}}
\end{figure}

The time-dependent analysis of the \CP asymmetry in $\Bd\to\jpsi\KS$ decays allows
a measurement of $\sin2\beta$. The latest result~\cite{Aaij:2015vza} from the LHCb collaboration
used 41560 flavour-tagged candidates (Figure~\ref{fig:jpsiKS} (a)) to measure the parameters
$S_{\jpsi\KS}$ = $+0.731 \pm 0.035 \pm 0.020$ and $C_{\jpsi\KS}$ = $-0.038 \pm 0.032 \pm 0.005$,
where $S_{\jpsi\KS} \approx \sin2\beta$ and the first uncertainty is of statistical origin and the second is systematic.
The correlation between the parameters is $\rho(S,C)$ = $0.483$.
Figure~\ref{fig:jpsiKS}(b) shows the signal asymmetry as a function of the decay time, where
the sinusoidal oscillation is clearly visible.
The dominant systematic uncertainty comes from knowledge of the tagging asymmetry of
the background events.  The result is consistent with the existing world average and of
similar precision to existing measurements, as shown in Figure~\ref{fig:global}(a).

\section{\phis\ from $\Bs\to\jpsi K^+K^-$ and $\Bs\to\jpsi\pi^+\pi^-$}

In the \Bs system the measurement of \phis\ can be performed via a
time-dependent analysis of $\Bs\to\jpsi \phi$ ($\phi\to K^+K^-$)
and $\Bs\to\jpsi\pi^+\pi^-$ decays. In these cases the final states
are ad-mixtures of both \CP-odd and \CP-even components such that
the decay angle information must also be used to disentangle the \CP
states. This gives rise to a rich structure and allows the simultaneous
determination of many \Bs mixing parameters from the $\Bs\to\jpsi \phi$
decay, such as $\phis, \dms, \Gs, \DGs, |\lambda_f|$. The LHCb 
collaboration has measured these parameters using 96\,000 
$\Bs\to\jpsi \phi$ decays~\cite{Aaij:2014zsa} with an effective tagging efficiency of
$\sim 3.0\%$, leading to $\phi_s  = -0.058  \pm  0.049  \pm  0.006$ rad
and $|\lambda| =  0.964  \pm  0.019 \pm 0.007$.
Only a $2\%$ contribution from the $K^+K^-$ S-wave is found
in a $60\mevcc$ window around the $\phi(1020)$ meson. The main
systematic uncertainties on the measurement of $\phi_s$ comes from
the description of the angular efficiency of the LHCb detector and 
event selection requirements.

\begin{figure}[t]
\centering{
\begin{overpic}[scale=0.3]{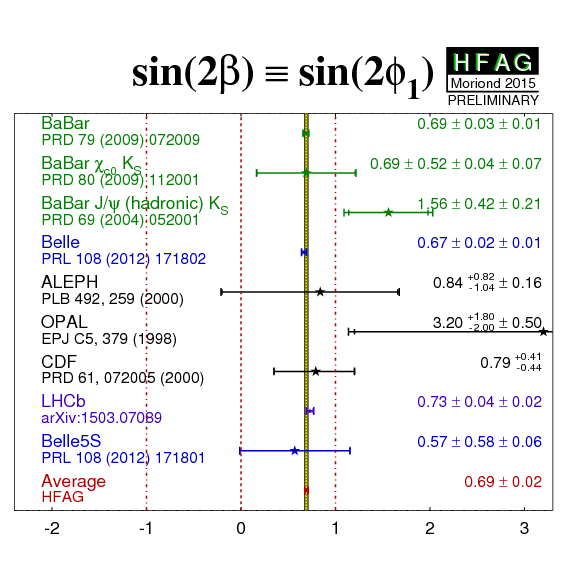}
 \put (0,85) {(a)}
\end{overpic}
\begin{overpic}[scale=0.67]{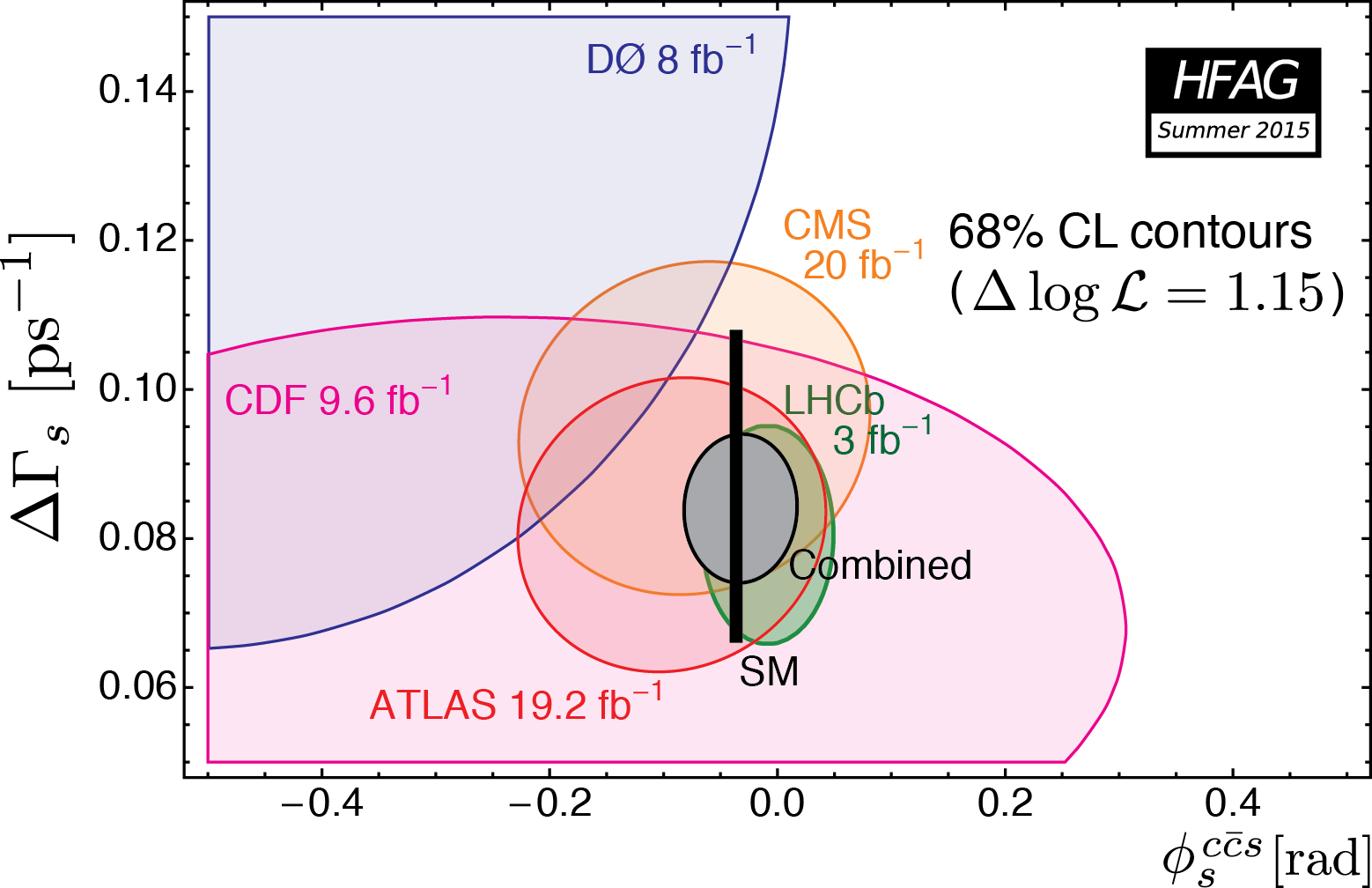}
 \put (20,55) {(b)}
\end{overpic}
}
\caption{Summary of measurements and the global fit result for (a) $\sin2\beta$ and (b) \phis\ and $\Delta\Gamma_s$~\cite{Amhis:2014hma}.
\label{fig:global}}
\end{figure}

In the case of $\Bs\to\jpsi\pi^+\pi^-$ a four-dimensional
amplitude analysis is performed to understand structure in the full $\pi^{+}\pi^{-}$ spectrum,
finding it to be $>97.7\%$ \CP-odd @ 95\% CL~\cite{Aaij:2014emv} with the main systematic 
uncertainty coming from the $\pi^+\pi^-$ resonance model. The measured
values of the \CP parameters from this analysis are~\cite{Aaij:2014dka}
$\phi_s  = 0.070  \pm  0.068  \pm  0.008$ rad
and $|\lambda| =  0.894  \pm  0.05 \pm 0.01$.
Assuming \CP violation in decay is the same in both the $\Kp\Km$ and $\pi^+\pi^-$ modes
is it possible to combining the measurements~\cite{Aaij:2014zsa} giving
the most precise determinations of the parameters 
$\phi_s   = -0.010  \pm  0.039$ rad and $|\lambda| =  0.957  \pm  0.017$, where
statistical and systematic uncertainties have been combined in quadrature.
Figure~\ref{fig:global}(b) shows the current status of all measurements of
$\phi_s$ and $\Delta\Gamma_s$ and the global fit, which is consistent with the
SM predictions~\cite{Charles:2015gya,Lenz:2006hd,Badin:2007bv},
indicating that any contributions from beyond-the-SM physics is a small.

\section{Controlling penguin pollution}

The contributions to $\phi_{d,s}$ from penguin decay topologies are
suppressed by $\epsilon = |V_{us}|^2/(1-|V_{us}|^2) \approx 0.05$ relative to the tree decay
and, as discussed in Section~\ref{sec:intro}, assumed to be negligible.
Given the precision with which $\phi_{d,s}$ are now known and the projected
sensitivity in the near-future it is it is
essential that such ``penguin pollution" is constrained
before trying to use these \CP violating phases in searches for new physics.
Since the
shift in $\phi_{d,s}$ from such contributions are difficult to calculate due to the 
non-perturbative nature of QCDit is crucial to
make experimental measurements of the size of the penguin contribution.
Various approaches exist to deal with this
problem~\cite{Fleischer:1999zi,Fleischer:2006rk,Faller:2008gt,DeBruyn:2014oga,Liu:2013nea,Frings:2015eva}.
One technique is to measure the \CP violating phase for different
polarisations of the final state, any observed differences then 
being interpreted as due to the pollution. This approach has been applied to the
analysis of $\Bs\to\jpsi\phi$ decays~\cite{Aaij:2014zsa} and no evidence of a
polarisation-dependent \CP violating phase is observed. The second is to study
decays where the penguin-to-tree ratio is not suppressed, as is discussed in more detail below.

\subsection{$\Bs \to J/\psi \overline{K}^{*}(892)^{0}$ and $\Bd \to J/\psi \rho^0$\label{sec:penguins}}

A suitable channel where the penguin amplitude is not suppressed relative to the tree is
$\Bs \to J/\psi \overline{K}^{*}(892)^{0}$~\cite{Fleischer:1999zi,Faller:2008gt,DeBruyn:2014oga}.
The amplitudes for this mode
and $\Bs \to J/\psi \phi$ are given by,
\begin{equation}
A(\Bs \to (\jpsi \overline{K^{*0}})_f) =  -\lambda {\cal A}_f \left[ 1 - a_f e^{i\theta_f} e^{i\gamma}\right], \ \ \ 
A(\Bs \to (\jpsi\phi)_f) =  (1-\lambda^2/2) {\cal A}^\prime_f \left[ 1 + \epsilon a^\prime_f e^{i\theta_f^\prime} e^{i\gamma}\right],
\end{equation}
where $f$ represents the polarisation of the final state, $\gamma$ is an angle of the CKM triangle
and ${\cal A}_f^{(\prime)}$
is a \CP-conserving hadronic matrix element. The parameters
$a_f^{(\prime)}$ and $\theta_f^{(\prime)}$ are the magnitude and phase of the penguin amplitude, respectively.
The first stage towards constraining the size of the penguin phase is to perform 
an angular analysis of the $\Bs \to J/\psi \overline{K}^{*0}$ decay products.
This allows the branching fraction of the decay to be measured along with the polarisation fractions
of the final state. By separating the sample into \Bs and \Bsb subsets, the \CP asymmetries for each polarisation
can also be measured after accounting for production and detection asymmetries~\cite{Aaij:2014bba,Aaij:2014nxa}.
The LHCb collaboration has recently published~\cite{Aaij:2015mea}
the results of this analysis using a sample of 1800 $\Bs \to J/\psi \overline{K}^{*0}$
signal events as seen in Figure~\ref{fig:psi2S}. The branching fraction is
measured to be ${\cal B}(\Bs\to\jpsi\overline{K^{*0}}) = ( 4.13 \pm 0.16 \pm 0.25 \pm 0.24\ (f_s/f_d))\times 10^{-5}$
and the \CP asymmetries are all consistent with zero at a precision of $\sim 10\%$, dominated
by the statistical uncertainty.

The second stage then relates these experimental observables
to the penguin parameters in both $\Bs \to J/\psi \overline{K}^{*0}$ and $\Bs \to J/\psi \phi$. 
A $\chi^2$ fit is performed to measure values for the penguin parameters
using the assumption of SU(3) flavour symmetry
($a_f^{\prime}=a_f$ and $\theta_f^{\prime} = \theta_f$),
a light-cone sum rule calculation~\cite{Straub:2015ica} of $|{\cal A}^\prime_f/{\cal A}_f|$ and
$\gamma = (3.2^{+6.3}_{-7.0})^\circ$~\cite{Charles:2015gya} . These are 
subsequently translated into values for the penguin phase shift that are shown to be 
consistent with zero, albeit with large uncertainties of $\sim 0.050$ rad.

\begin{figure}[t]
\begin{minipage}{0.46\linewidth}
\centering{
	\includegraphics[width=1.1\linewidth]{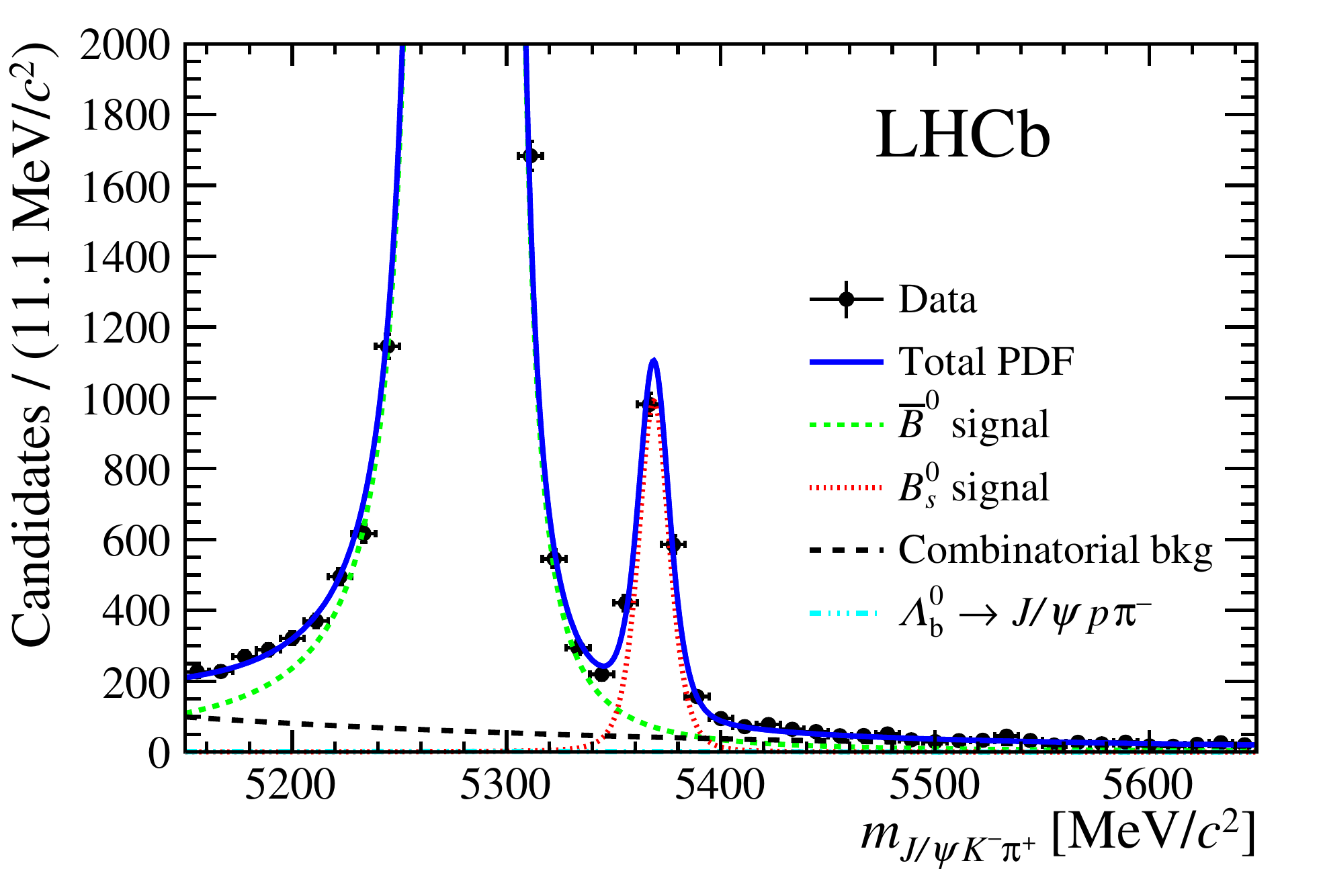}
	}
\caption{Distribution of $J/\psi \overline{K}^{*0}$ invariant mass for all $\Bsb \to J/\psi \overline{K}^{*0}$ candidates.
\label{fig:psi2S}}
\vspace{1.05cm}
\end{minipage}
\hfill
\begin{minipage}{0.51\linewidth}
\centering{
	\includegraphics[width=\linewidth]{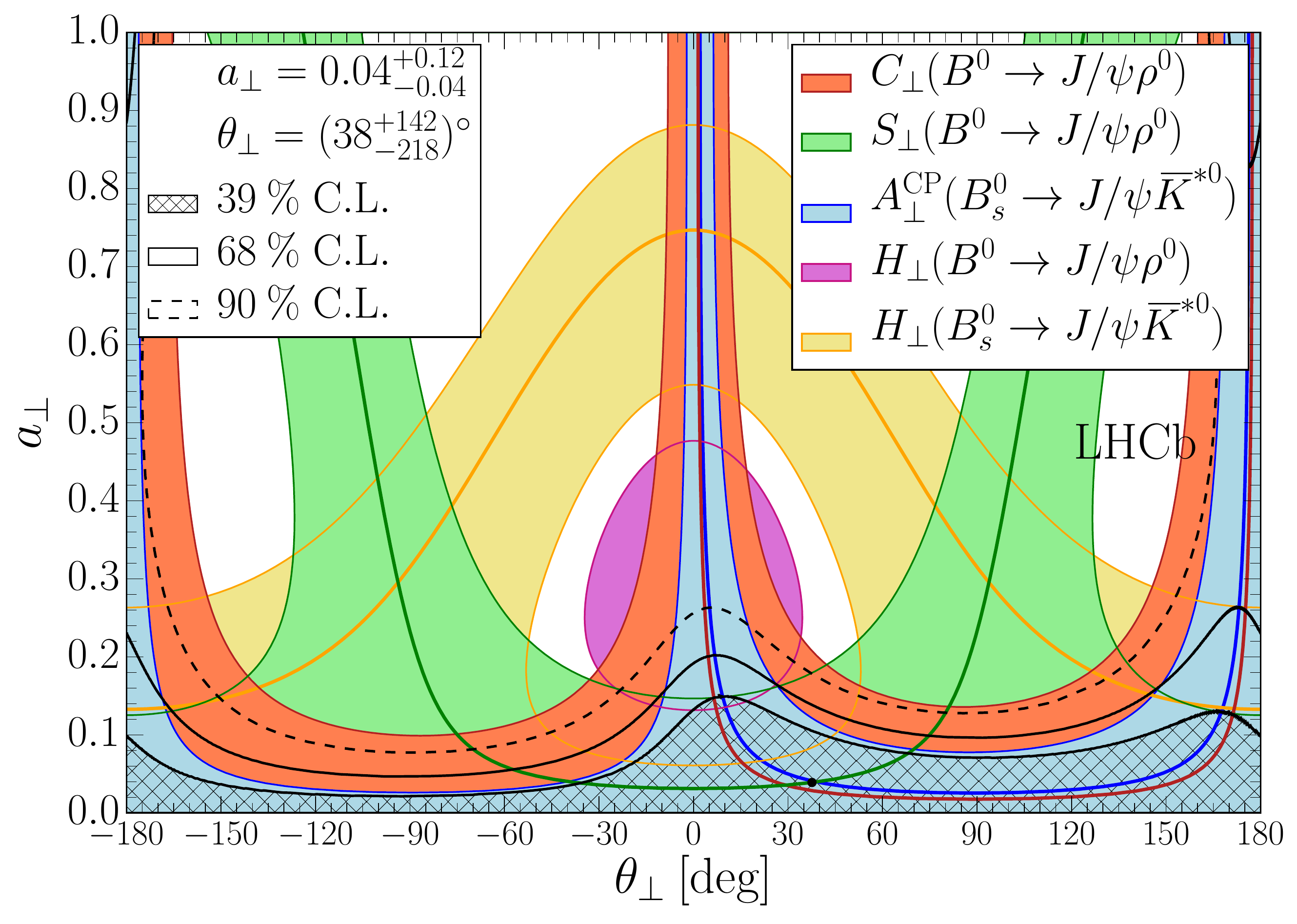}
	}
\caption{Limits on the penguin parameters $a_\perp$ and $\theta_\perp$ obtained from intersecting contours
derived from \CP asymmetries and branching fraction measurements in $\Bs \to J/\psi \overline{K}^{*0}$ and $\Bd \to J/\psi \rho^0$.
Superimposed are  the confidence-level contours from a $\chi^2$ fit to the data.
\label{fig:penguin_shift}}
\end{minipage}
\end{figure}

The dominate systematic 
uncertainty from the theoretical calculation of $|{\cal A}^\prime_f/{\cal A}_f|$
can be removed and the overall statistical uncertainty reduced by including
complementary information on the \CP violating phases measured by
the LHCb collaboration in each polarisation state of the $\Bd \to J/\psi \rho^0$
($\rho^0\to\pi^+\pi^-)$ channel~\cite{Aaij:2014vda}.
Figure~\ref{fig:penguin_shift} shows the constraints from each measurement for one polarisation state. 
The central values of the parameters are measured using a $\chi^2$ fit, the result of which is shown
by the black contours in Figure~\ref{fig:penguin_shift}.
Again, the penguin phase shifts are found to be consistent with zero, but now
determined to a precision of $\sim 0.015$ rad through the additional 
information from the $\Bd \to J/\psi \rho^0$ channel. This should be compared
with the current experimental precision on $\phi_{d,s}$ that are at $\sim 0.030$ rad.

\subsection{$\Bs \to J/\psi \KS$}

\begin{figure}[t]
\centering{
\begin{overpic}[width=0.45\linewidth]{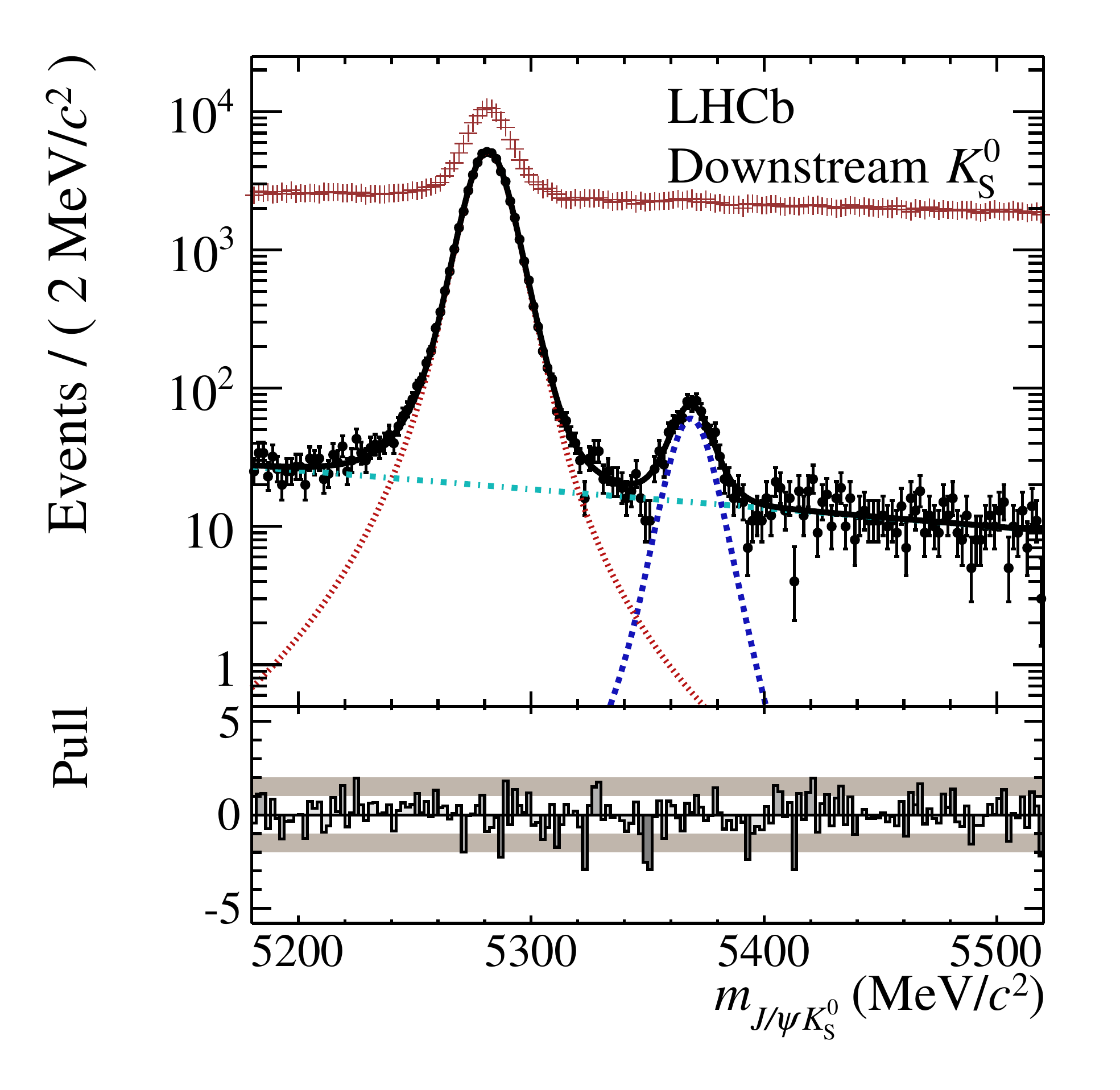}
 \put (60,70) {(a)}
\end{overpic}
\begin{overpic}[width=0.45\linewidth]{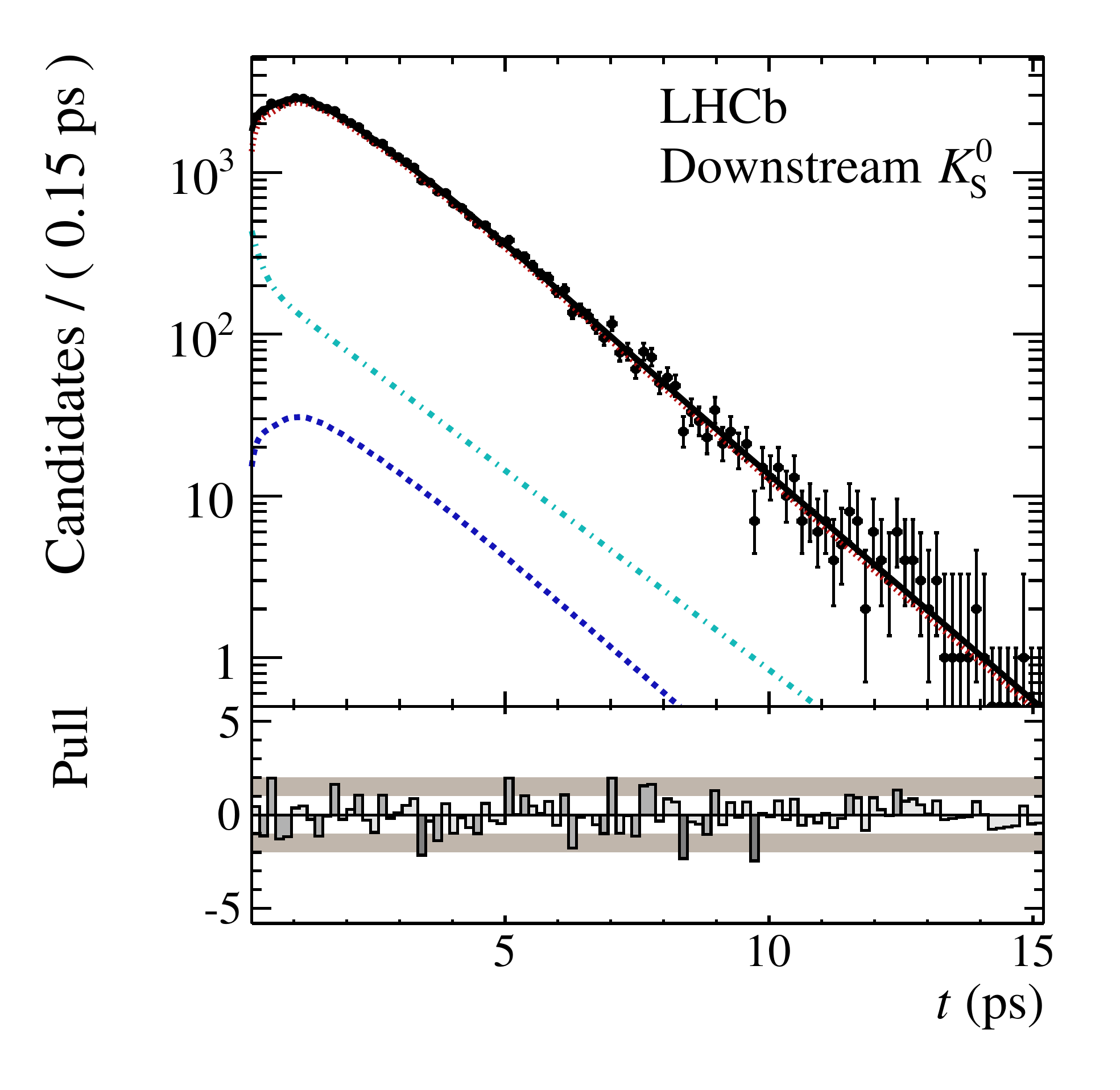}
 \put (60,70) {(b)}
\end{overpic}
}
\caption{Distributions of (a) $J/\psi \KS$ invariant mass and (b) decay time of $\Bs \to J/\psi \KS$ candidates. In (a)
the distribution before (red) and after (black) the machine learning selection is applied are shown.
\label{fig:Bs2JpsiKS}}
\end{figure}

The golden mode for controlling the penguin phase shift to $\phi_d$  is $\Bs \to J/\psi \KS$.
This is related to $\Bd\to\Jpsi\KS$ via the U-spin symmetry of strong
interactions~\cite{Fleischer:1999zi} but is CKM 
suppressed making it more difficult to observe the decay. The LHCb collaboration has
recently~\cite{Aaij:2015tza} used machine learning techniques to suppress backgrounds
from combinatorics and mis-reconstructed $\Bd\to\jpsi\Kst$ decays to observe $\Bs \to J/\psi \KS$.
Figure~\ref{fig:Bs2JpsiKS} shows the distribution of invariant mass and \Bs decay time
for a subset of $\Bs \to J/\psi \KS$ candidates. In total $\sim 900$ signal events are found
with an effective tagging efficiency of $\sim 4\%$. These are used to measure the decay time
dependent \CP violation observables $A_{\Delta\Gamma}, C, S$ defined in Equation~(\ref{eqn:asymm}). 
All are found to be consistent with zero. The large statistical uncertainty on these measurements
is insufficient to constrain the size of the penguin shift to $\phi_d$ but this proof-of-principle
measurement indicates what is possible with future datasets collected by LHCb.

\subsection{$\Bsb \to \psi(2S) K^+\pi^-$}

The LHCb collaboration has recently observed~\cite{Aaij:2015wza} the $\Bsb \to \psi(2S) K^+\pi^-$
decay, measuring the ratio of branching ratios
${\cal B}(\Bsb \to \psi(2S) K^+\pi^-)/{\cal B}(\Bsb \to \psi(2S) K^+\pi^-) = 5.38 \pm 0.36\ \stat \pm 0.22\ \syst \pm 0.31\ (f_s/f_d) \%$.
Figure~\ref{fig:psi2S} shows the distributions of $\psi(2S) K^+\pi^-$ and $K^+\pi^-$ invariant masses from the sample.
Using a four-dimensional amplitude analysis the fraction of decays proceeding via an intermediate $K^*(892)^0$
meson is measured to be $0.645 \pm 0.049\ \stat \pm 0.049\ \syst$ and its longitudinal polarisation fraction,
$f_{\rm 0} = 0.524 \pm 0.056\ \stat \pm 0.029\ \syst$. No exotic $Z^+ \to \psi(2S)\pi^+$ component~\cite{Aaij:2014jqa} was
observed with the current data sample but this mode could prove useful in the future to help understand
the nature of the exotic charmonium states. In addition, using the same technique as discussed in
Section~\ref{sec:penguins}, this mode could help to constrain the size of penguin pollution to the
\CP violating phase in $\Bs \to \psi(2S) \phi$ decays.

\begin{figure}[t]
\centering{
\begin{overpic}[scale=0.38, clip=true, trim=16mm 28mm 0mm 0mm]{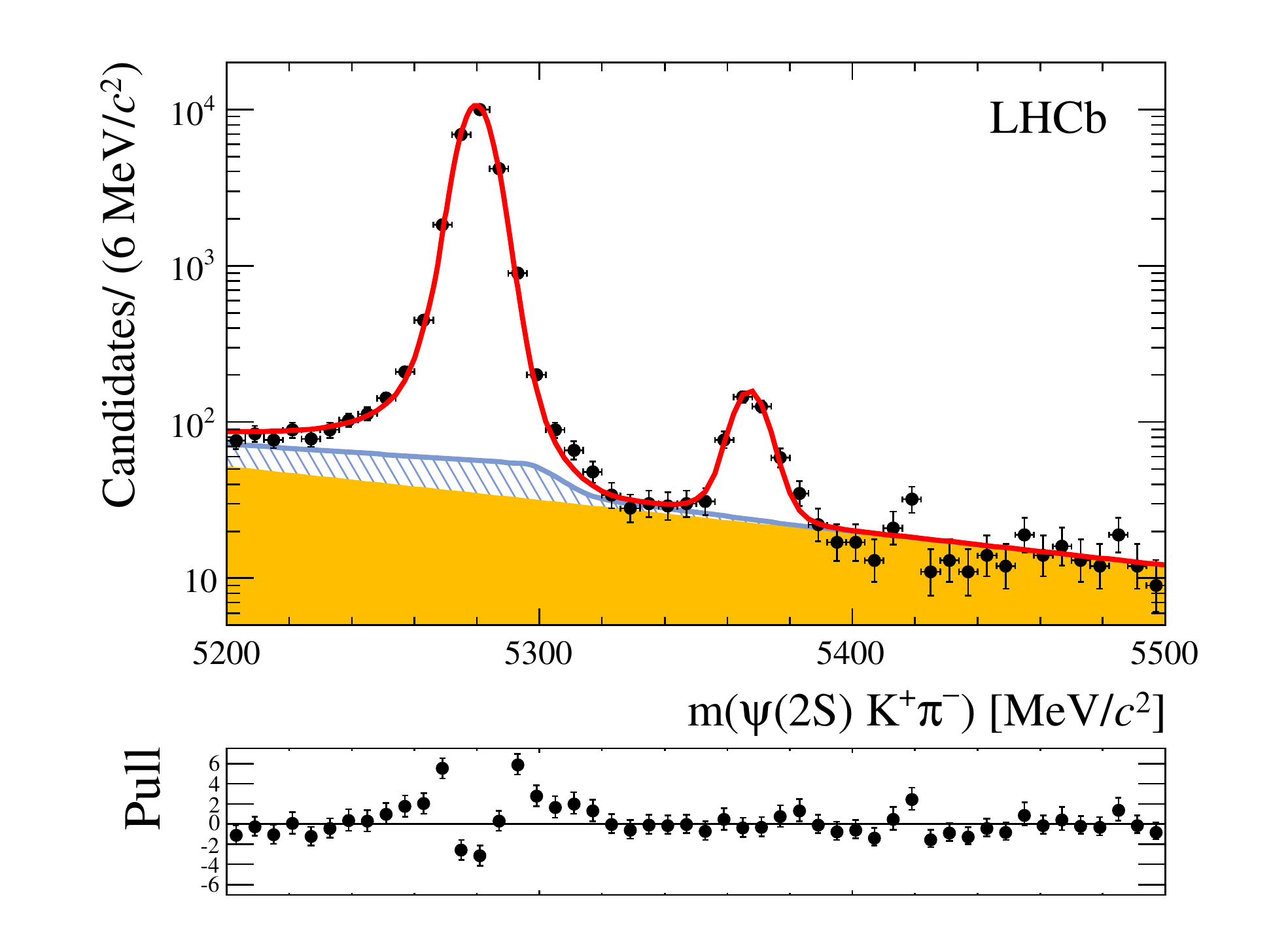}
 \put (50,40) {(a)}
\end{overpic}
\begin{overpic}[scale=0.38, clip=true, trim=16mm 40mm 0mm 0mm]{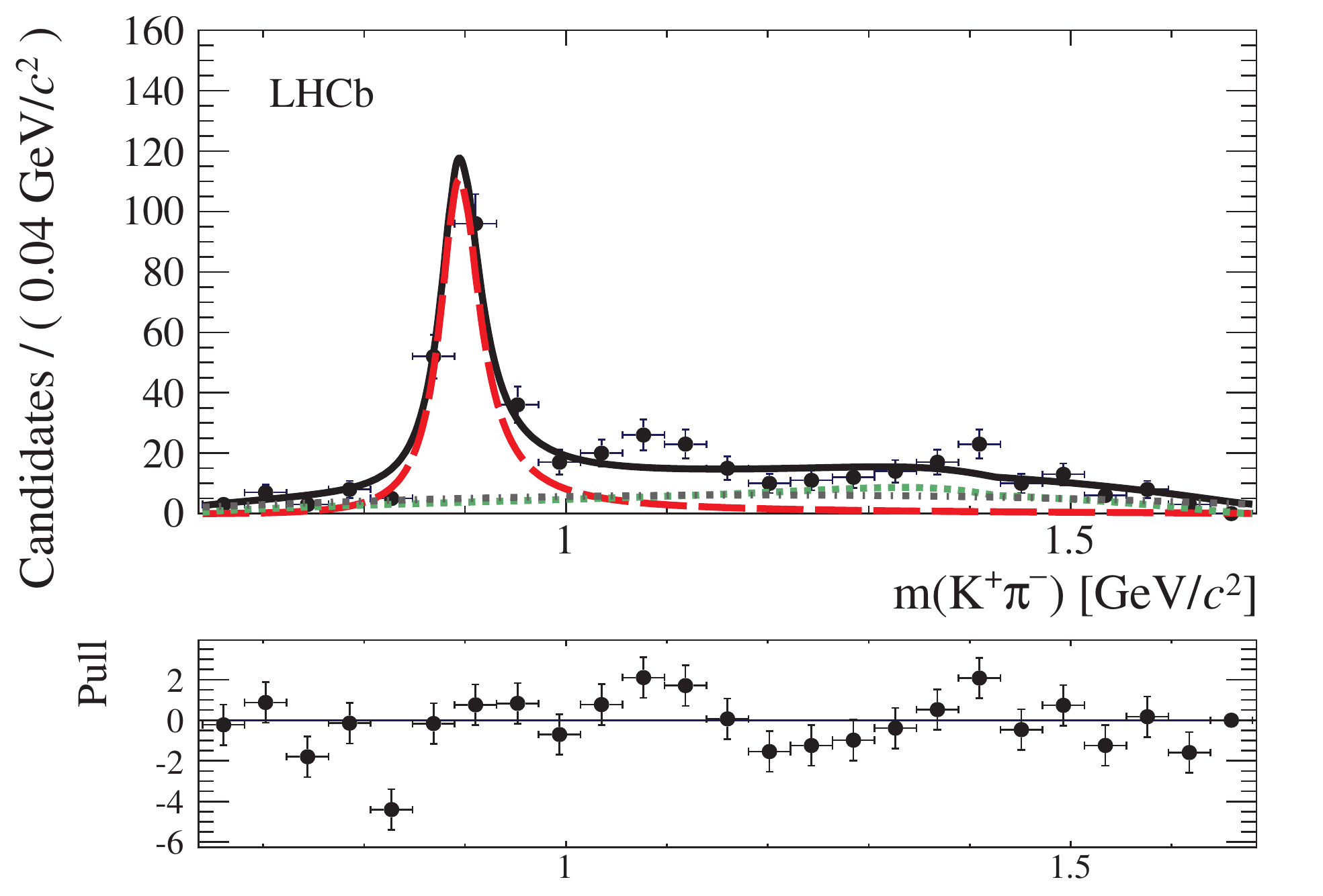}
 \put (50,40) {(b)}
\end{overpic}
}
\caption{Distributions of (a) $\psi(2s) K^+\pi^-$  and (b) $K^+\pi^-$ invariant masses of $\Bsb \to \psi(2s) K^+\pi^-$ candidates.
In (b) the contributions from the $K^*(892)^0$ (red-dashed), NR (green-dotted) and background (grey-dashed-dotted) components
are visible.
\label{fig:psi2S}}
\end{figure}

\section{Summary}

These proceedings have presented the latest measurements of \CP violation in \B meson
mixing using $\B\to\jpsi X$ decays collected by the LHCb experiment during $pp$ collision runs at the LHC
at $\sqrt{s} = 7$ and $8 \tev$. These represent the most precise determination of the
\CP violating phase $\phi_s$ using $\Bs\to\jpsi\phi$ and $\Bs\to\jpsi\pi^+\pi^-$ decays
along with a very competitive measurement of the corresponding phase in the \Bd system using
$\Bd\to\jpsi\KS$ decays. With this improved precision it is now essential that contributions
to these decay processes from suppressed, but hard-to-calculate, penguin diagrams are under control. By
making use of \B meson decays where the penguin decay processes are not suppressed relative 
to the tree-level contribution (such as $\Bsb\to\jpsi K^{*0}$, $\Bd\to\jpsi\rho^0$ and $\Bs\to\jpsi\KS$)
the LHCb collaboration has been able to constrain the size of the so-called penguin pollution, finding it to
be small. The collaboration looks forward to collecting even more data during Run-2 of the LHC and
beyond, where making further precision measurements of \CP violating parameters will be 
essential as we search for signs of beyond-the-SM physics.

\section*{Acknowledgements}

The author would like to thank Alex Lenz for commenting on the text.

\end{document}